\documentclass[%
 reprint,
 amsmath,amssymb,
 aps,
]{revtex4-1}
\pdfoutput=1
\usepackage{graphicx}
\usepackage{dcolumn}
\usepackage{bm}
\usepackage{multirow}
\usepackage{array}
\usepackage{xcolor,soul}
\usepackage{booktabs}
\usepackage{underscore}

\def\REF#1{\par\hangindent\parindent\indent\llap{#1\enspace}\ignorespaces}
\newcommand{\ucite}[1]{$^{[#1]}$}
\renewcommand{\cite}[1]{\,[#1]}
\newcommand{\etal}{\textit{et~al}.\ }

\begin{document}

\title{Magneto-elastic coupling in a sinusoidal modulated magnet Cr$_2$GaN}
\thanks{This work is supported by the National Natural Science Foundation of China (Grant Nos. 11822411, 12061130200, 11961160699, 11974392 and 52101236), the National Key Research and Development Program of China (Grant Nos. 2018YFA0704200, 2017YFA0303100 and 2020YFA0406003), the Strategic Priority Research Program (B) of the CAS (Grant No. XDB25000000), K. C. Wong Education Foundation (GJTD-2020-01), the Youth Innovation Promotion Association of the CAS (Grant No. Y202001), Beijing Natural Science Foundation (Grant No. JQ19002), and the Newton Advanced Fellowship funding from the Royal Society of UK (Grant No. NAF_R1_201248).This work is based on neutron diffraction experiments performed at OSIRIS (Proposal No. RB1610019), ISIS Facility, Rutherford Appleton Laboratory, UK and WOMBAT (Proposal No. P6014), Australian Centre for Neutron Scattering, Australian Nuclear Science and Technology Organisation, Australia.}

\author{Hui-Can Mao $^{1,2}$, Yu-Feng Li$^{3}$, Qing-Yong Ren$^{4,5}$, Mi-Hai Chu$^{6}$, Helen E. Maynard-Casely$^{7}$, Franz Demmel$^{8}$, Devashibhai Adroja$^{8}$, Hai-Hu Wen$^{3}$, Yin-Guo Xiao$^{6}$$^{\dag}$  and Hui-Qian Luo$^{1,9}$ }

\thanks{Corresponding authors. Email: y.xiao@pku.edu.cn; hqluo@iphy.ac.cn}
\affiliation{
$^{1}${Beijing National Laboratory for Condensed Matter Physics, Institute of Physics, Chinese Academy of Sciences, Beijing 100190, China}\\
$^{2}${University of Chinese Academy of Sciences, Beijing 100049, China}\\
$^{3}${National Laboratory of Solid State Microstructures and Department of Physics, Collaborative Innovation Center of Advanced Microstructures, Nanjing University, Nanjing 210093, China}\\
$^{4}${Institute of High Energy Physics, Chinese Academy of Sciences, Beijing 100049, China}\\
$^{5}${Spallation Neutron Source Science Center, Dongguan 523803, China}\\
$^{6}${School of Advanced Materials, Peking University, Shenzhen Graduate School, Shenzhen 518055, China}\\
$^{7}${Australian Centre for Neutron Scattering, Australian Nuclear Science and Technology Organisation, Lucas Heights NSW-2232, Australia}\\
$^{8}${ISIS Facility, Rutherford Appleton Laboratory, Chilton, Didcot Oxon OX11 0QX, United Kingdom}\\
$^{9}${Songshan Lake Materials Laboratory, Dongguan, Guangdong 523808, China}
}

\begin{abstract}
We use neutron powder diffraction to investigate the magnetic and crystalline structure of Cr$_2$GaN. A magnetic phase transition is identified at $T \approx 170$ K, whereas no trace of structural transition is observed down to 6 K. Combining Rietveld refinement with irreducible representations, the spin configuration of Cr ions in Cr$_2$GaN is depicted as an incommensurate sinusoidal modulated structure characterized by a propagating vector ${k}$=(0.365, 0.365, 0). Upon warming up to the paramagnetic state, the magnetic order parameter closely resembles  to the temperature dependence of $c$-axis lattice parameter, suggesting strong magneto-elastic coupling in this compound. Therefore, Cr$_2$GaN provides a potential platform for the exploration of magnetically tuned properties such as magnetoelectric, magnetostrictive and magnetocaloric effects, as well as their applications.
\end{abstract}

\pacs{Valid PACS appear here}
\maketitle


The M$_{n+1}$AX$_n$ (MAX) phases, crystallized in hexagonal P6$_3$/mmc, where M is a transition metal element, A is an IIIA or IVA group element, X is either C or N and n varies from 1 to 3, have attracted considerable interest for a unique combination of metallic properties such as good electrical conductivity, and ceramic properties like high elastic moduli \ucite{1-3}. The MAX compounds are composed of near-close-packed layers of M$_6$X octahedra interleaved with layers of pure A atoms. These MAX phases are extremely resistant to oxidation and thermal shock, elastically stiff, while at the same time they exhibit high electrical and thermal conductivity which are machinable \ucite{4,5}. In addition to the unique metallic and ceramic characters in inherently nanolaminated MAX phases, magnetism characterized by high stability and tunable anisotropic properties are fascinating for both fundamental materials science and practical applications, such as spintronics manipulated by the interlayer exchange coupling, and metallic magnetism realized in the geometrically frustrated two dimensional plane \ucite{6-10}.

Mostly magnetic MAX phases reported to date, either in bulk or thin film form, are composed of Cr or Mn as transition metal elements \ucite{6}. The significantly correlated nature of the $d$ orbitals of Cr and Mn in MAX compounds generates diverse magnetic interactions under either ferromagnetic (FM) or antiferromagnetic (AFM) order, which strongly depend on the chemical composition and other tuning parameters like temperature or pressure. For example, Cr$_2$GeC is probably located in the vicinity of a FM quantum critical point, whereas the Mn doped Cr$_2$GeC shows itinerant-electron ferromagnetism in a wide range of the degree of electron localization \ucite{11}.  Interestingly, Mn$_2$GaC undergoes a magnetic transition ($\sim$ 210 K) from a collinear AFM state to a non-collinear AFM configuration due to the spin reorientation, accompanied by a $c$-axis lattice compression with decreasing temperature \ucite{7,8}. This suggests a strong coupling between spin and lattice, so called ``magneto-eleastic" coupling, in the layered MAX phase. Such rich magnetic properties in MAX phases open a pathway to explore new functionalities for spintronic, magnetoelectric and magnetocaloric applications by enabling the manipulation of electronic, magnetic and crystalline structure.

Analogously, a spin-density-wave (SDW) transition has been found in MAX phase Cr$_2$GaN at T$_N$ = 170 K by measuring the zero field muon spin relaxation (ZF-$\mu$SR), magnetic susceptibility and resistivity \ucite{12,13}. Furthermore, nuclear magnetic resonance (NMR) experiments suggest that the static and distributed internal field is expected for the frustrated magnetism or incommensurate SDW state\ucite{13} in Cr$_2$GaN. Such rich and complicated magnetic properties in Cr$_2$GaN will inspire us to determine its magnetic ground state and further explore the interactions between spin and lattice by neutron diffraction experiments.

\begin{figure}
\includegraphics[width=3.4in]{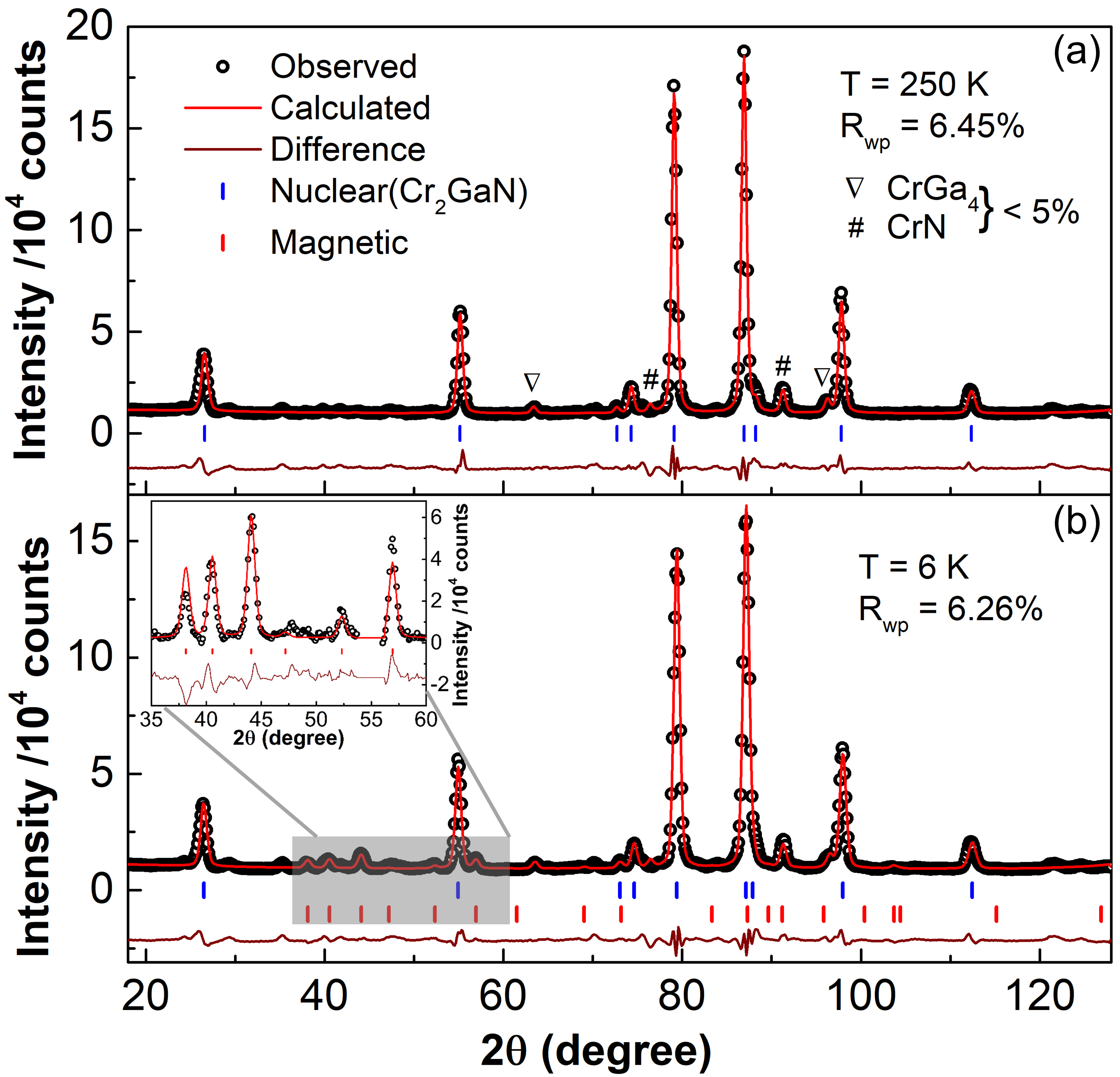}
\caption{\label{fig:wide}The rietveld refinement results with hexagonal phase P6$_3$/mmc at (a) 250 K and (b) 6 K. The vertical blue lines represent the positions of the bragg peaks in Cr$_2$GaN, red lines designate that of magnetic peaks, inverted triangle is the impurities of CrGa$_4$ and CrN. The inset in Fig.1b gives a schematic show of the refinement of magnetic phase in the range of 2${\theta}$ $\sim$ 35$^{\circ}$ - 60$^{\circ}$.}
\end{figure}

The polycrystalline samples of Cr$_2$GaN were synthesized by the solid state reaction method as described in Reference \ucite{12}. Neutron powder diffraction experiments were carried out on both OSIRIS high-resolution diffractometer at ISIS facility, STFC Rutherford Appleton Laboratory (Proposal No. RB1610019) and WOMBAT high-intensity diffractometer at Australian Nuclear Science and Technology Organisation (Proposal No. P6014) with wavelength $\lambda$ = 2.95 \AA. To obtain diffraction data with good counting statistics and homogeneous background, about 5 g of Cr$_2$GaN powders were laid flat between two layers of aluminum foil, then rolled up and placed against the wall of Aluminum can for the experiment at OSIRIS diffractometer, whereas the same batch of samples $\sim$ 5 g were sealed in a vanadium can at WOMBAT diffractometer. The full pattern measurements were conducted on WOMBAT diffractometer at 6 K and 250 K by covering the scattering angle $2\theta$ from 15 to 136 degrees, whereas the temperature-dependent diffraction data ranging from 10 K to 250 K were collected at OSIRIS diffractometer with interplanar $d$-spacing from  3.1 to 5.3 \AA, including a nuclear reflection (004) and several magnetic reflections. For the structural refinement, we focus on the WOMBAT data by using Rietveld method and irreducible representation theory embedded in the FullProf \ucite{14,15} and SARAH \ucite{16} software package.

\begin{table}
\caption{\label{tab:table1}Crystallographic data of Cr$_2$GaN at 250 K and 6 K.}
\begin{ruledtabular}
\begin{tabular}{ccccccccc}
&&        &250 K     && 6 K   \\
\hline
&space group                 && P6$_3$/mmc        &&  P6$_3$/mmc \\
 &$a$(\AA)                   && 2.861(5)          && 2.850(1)  \\
 &$c$(\AA)                   && 12.671(9)         && 12.713(1)  \\
 &z$_{Cr}$                   && 0.087(2)          && 0.089(1)  \\
 &B$_{iso}$(Cr)(\AA$^{2}$)   && 1.118(2)          && 0.350(1)  \\
 &B$_{iso}$(Ga)(\AA$^{2}$)   && 0.839(2)          && 0.515(8) \\
 &B$_{iso}$(N)(\AA$^{2}$)    && 1.112(6)          && 0.960(8) \\
\hline
&atom  &Wyckoff   &x      &y     &z         \\
&Cr     &4f            &1/3    &2/3  &z$_{Cr}$  \\
&Ga    &2d           &1/3    &2/3  &0.75      \\
&N      &2a           &0       &0     &0          \\
\end{tabular}
\end{ruledtabular}
\end{table}

The raw data of neutron diffraction patterns at 250 K and 6 K obtained from WOMBAT diffractometer are presented in Fig.1. To quantitatively analyze the crystal structure and magnetic structure of Cr$_2$GaN, we have performed the Rietveld refinement for these two data sets. The majority of Bragg reflections can be indexed by a hexagonal phase of Cr$_2$GaN with space group P6$_3$/mmc for both patterns at 250 K and 6 K, whereas some weak reflections indicate the existence of very small amounts of CrGa$_4$ and CrN impurity phases (\textless  5 \%) [Fig.1(a)]. The parameters for the quality of this two fittings, including profile factor R$p$ and weighted profile factor R$wp$, are 4.75(3), 6.44(5) at 250 K and 4.88(4), 6.45(3) at 6 K, respectively. All crystallographic parameters listed in Table I are mostly consistent with previous X-ray diffraction results \ucite{12}. Since both data sets at 250 K and 6 K can be fully identified with the same space group P6$_3$/mmc, we thus conclude that there is no observable structural phase transition in Cr$_2$GaN down to 6 K. As illustrated in Fig.1(b), a few magnetic peaks emerge at low temperature of 6 K ranging from 35$^{\circ}$ to 60$^{\circ}$, indicating the formation of a long-rang magnetic order in Cr$_2$GaN. All magnetic peaks at 6 K can be well indexed with a propagation vector  ${k}$=(0.365,0.365,0), suggesting an incommensurate magnetic structure.

\begin{figure*}
\includegraphics[width=5in]{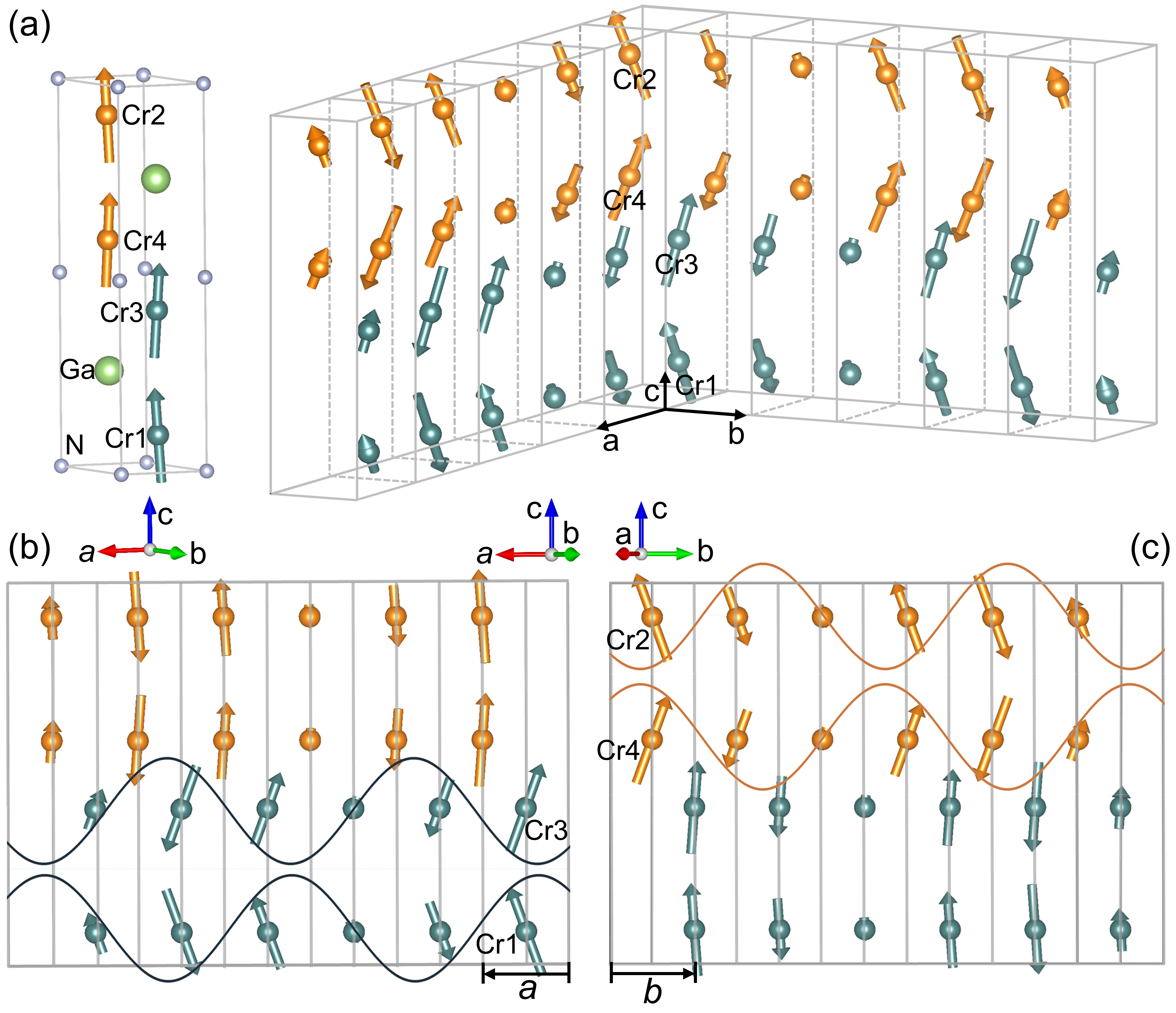}
\caption{\label{fig:wide}(a)Three-dimensional sinusoidal modulated magnetic structure of Cr$_2$GaN at 6 K in single unit cell and supercell. Two-dimensional sinusoidal modulated magnetic structure of Cr$_2$GaN at 6 K along (b) $a$-axis and (c) $b$-axis.}
\end{figure*}

The magnetic symmetry analysis for Cr$_2$GaN are performed using the program SARAH, by which the allowed symmetry coupling in the form of irreducible representation (IR) and their basis vectors (BVs) can be obtained. With the incommensurate magnetic propagation vector $k$ = (0.365, 0.365, 0), the Cr ion occupying at crystallographic Wyckoff site (4f) has four one-dimensional irreducible magnetic representations including $\Gamma$$_1$, $\Gamma$$_2$, $\Gamma$$_3$ and $\Gamma$$_4$, and each of them has three basis vectors with only real components. To determine the magnetic structure of Cr$_2$GaN, we use all possible IRs in the magnetic structural model to fit the neutron diffraction pattern at 6 K.

\begin{table}
\caption{\label{tab:table2}Nonzero basis vectors (BVs) of the irreducible representations (IRs) and positional coordinates for Cr atoms that are used to describe the sinusoidal modulated magnetic structure of Cr$_2$GaN with space group P6$_3$/mmc and propagation vector $k$=(0.365, 0.365, 0).}
\begin{ruledtabular}
\begin{tabular}{cccccccccccccc}

 &&&\multicolumn{3}{c}{BV components}   &\multicolumn{8}{c}{Positional coordinates}\\
   \cline{4-6} \cline{7-14}
  \rule{0pt}{11pt}
 IR &BV &Atom &{$m$$_{||a}$} &{$m$$_{||b}$} &{$m$$_{||c}$} &&&$x$ &&&&$y$ &$z$ \\
\colrule

$\Gamma$$_3$ &$\psi$$_7$ &1 &1  &0  &0 &&&$x$           &&&&$y$           &$z$             \\
&                        &2 &0  &-1 &0 &&&${\bar{x}}$+1 &&&&${\bar{y}}$+1 &${\bar{z}}$+1    \\
&                        &3 &-1 &0  &0 &&&$x$           &&&&$y$           &${\bar{z}}$+$1/2$ \\
&                        &4 &0  &1  &0 &&&${\bar{x}}$+1 &&&&${\bar{y}}$+1 &$z$+$1/2$          \\
             &$\psi$$_8$ &1 &0  &1  &0 &&&$x$           &&&&$y$           &$z$             \\
&                        &2 &-1 &0  &0 &&&${\bar{x}}$+1 &&&&${\bar{y}}$+1 &${\bar{z}}$+1    \\
&                        &3 &0  &-1 &0 &&&$x$           &&&&$y$           &${\bar{z}}$+$1/2$ \\
&                        &4 &1  &0  &0 &&&${\bar{x}}$+1 &&&&${\bar{y}}$+1 &$z$+$1/2$          \\
             &$\psi$$_9$ &1 &0  &0  &1 &&&$x$           &&&&$y$           &$z$             \\
&                        &2 &0  &0  &1 &&&${\bar{x}}$+1 &&&&${\bar{y}}$+1 &${\bar{z}}$+1    \\
&                        &3 &0  &0  &1 &&&$x$           &&&&$y$           &${\bar{z}}$+$1/2$ \\
&                        &4 &0  &0  &1 &&&${\bar{x}}$+1 &&&&${\bar{y}}$+1 &$z$+$1/2$          \\

\end{tabular}
\end{ruledtabular}
\end{table}

We find that IR $\Gamma$$_3$ is the best fitting of all magnetic peaks, whereas the other three representations can be excluded due to poor fits of magnetic peaks. The decomposition of the magnetic representation $\Gamma$$_3$ is given in Table II. According to $\Gamma$$_3$, the Cr moments are ordered in a sinusoidal structure in the magnetic unit cell and the refined vector components of the magnetic moment are $m$$_x$ = 0.4 ${\mu}$$_B$, $m$$_y$ = -0.05 ${\mu}$$_B$ and $m$$_z$ = 1.5 ${\mu}$$_B$. The Cr ion at 4f Wyckoff site has 4 equivalent Cr ions in the unit cell marked with Cr1(0.333, 0.667, 0.089), Cr2(0.667, 0.333, 0.911), Cr3(0.333, 0.667, 0.411) and Cr4(0.667, 0.333, 0.589) in Fig.2(a), and their ($m$$_x$, $m$$_y$, $m$$_z$) are (0.4, -0.05, 1.5)${\mu}$$_B$, (0.05, -0.4, 1.5)${\mu}$$_B$, (-0.4, 0.05, 1.5)${\mu}$$_B$ and (-0.05, 0.4, 1.5)${\mu}$$_B$ respectively. It is clear that the magnetic moment components of Cr ions are lying in $a$-$c$ and $b$-$c$ planes. More precisely, we note that Cr1 and Cr3 moments in $a$-$c$ plane (or Cr2 and Cr4 in $b$-$c$ plane) form a sinusoidal wave modulated structure model along $a$-axis (or $b$-axis) as shown in Fig.2(b) and Fig.2(c), respectively. The estimated value of the total ordered moment is about 1.56 ${\mu}$$_B$, which is slightly larger than the reported results of ${\mu}$SR experiments ($< 1$ $\mu_B$ ) \ucite{13}.

\begin{figure*}
\includegraphics[width=4.5in]{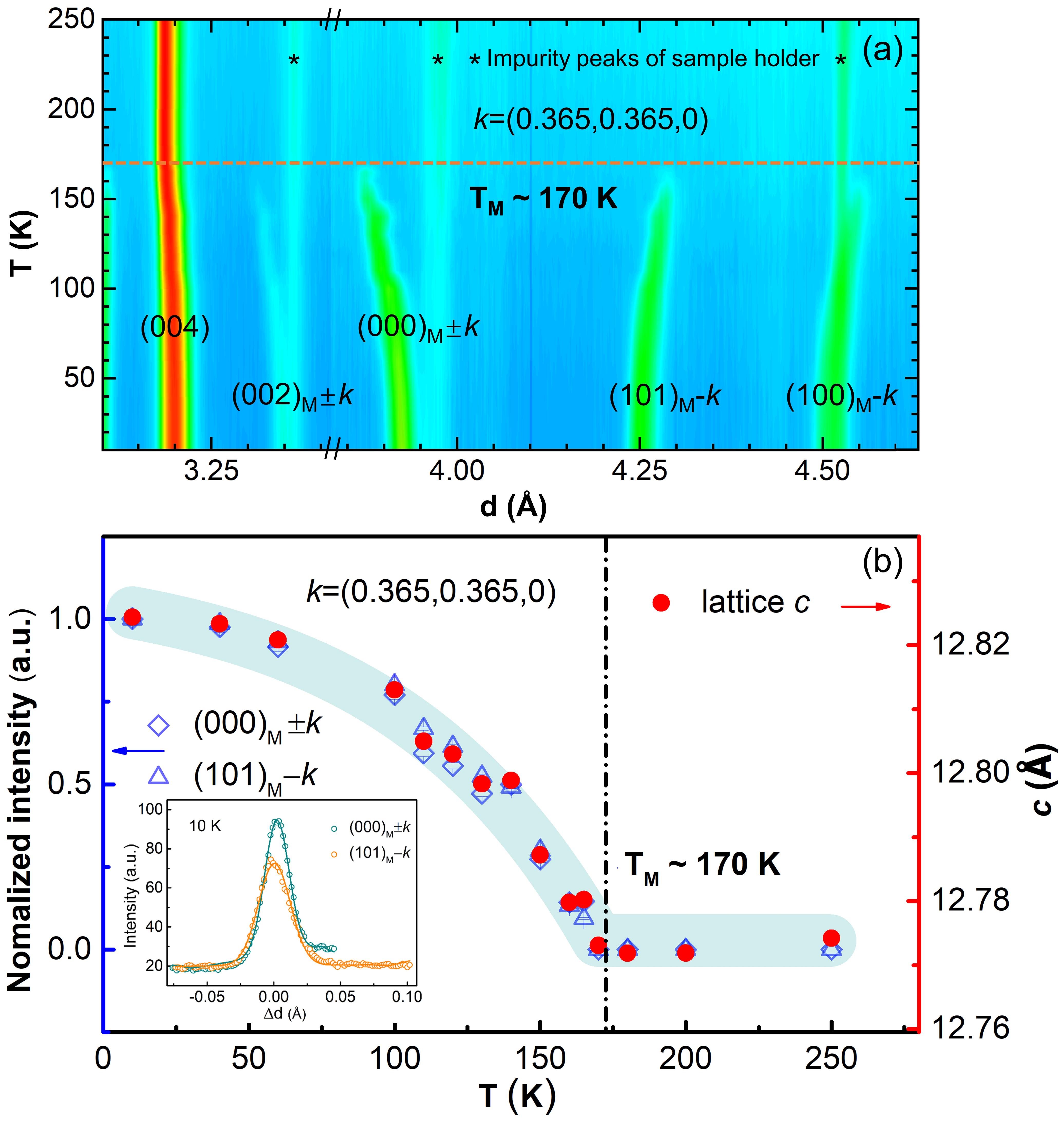}
\caption{\label{fig:wide}(a) The temperature-dependent nuclear peak (004) and magnetic peaks of (002)$_M$$\pm$$k$, (000)$_M$$\pm$$k$, (101)$_M$-$k$ and (100)$_M$-$k$. (b) Temperature dependence of the normalized intensity of magnetic peaks (000)$_M$$\pm$$k$ and (101)$_M$-$k$ and lattice constant $c$. The inset in Fig.3b is the fitting of magnetic peaks at 10 K. For these magnetic peaks, the fits to Gauss function are performed for different temperatures, and the obtained area represent their intensity of magnetic peaks.}
\end{figure*}

Now we turn to the diffraction pattern covering interplanar distance d $\sim$ (3.1 - 4.65 \AA) as shown in Fig.3, which were obtained from OSIRIS diffractometer with measured temperatures range from 10 K to 250 K. In Fig.3(a), the four magnetic peaks (002)$_M$$\pm$$k$, (000)$_M$$\pm$$k$, (101)$_M$-$k$ and (100)$_M$-$k$, centered around 3.32 \AA, 3.85 \AA, 4.28 \AA\ and 4.48 \AA, gradually become weaker with increasing temperature and disappear completely above magnetic phase transition temperature $T_M\approx$ 170 K. Notably, the positions of all magnetic peaks shift upon warming up to $T_M$, so does the position of (004) nuclear peak. As both the magnetic and crystalline structures keep the same below $T_M$, the peak shift of (004) toward smaller $d$ indicates the lattice parameter $c$ also decreases. The lattice parameter $c$ in Fig.3b obtained from OSIRIS diffractometer are slightly larger than that in Table I obtained from WOMBAT diffractometer, which are mainly attributed to the difference between the two diffractometers. The order parameter of the sinusoidal magnetism can be represented by the normalized intensities of two magnetic peaks (101)$_M$-$k$ and (000)$_M$$\pm$$k$, which are plotted together with the lattice constant $c$ in Fig.3(b) for comparison. Surprisingly, all these parameters decrease simultaneously in a uniform manner with increasing temperature until a kink occurs at 170 K, in perfect agreement with the transition temperature reported previously \ucite{12,13}. The analogous temperature-dependent behavior in the wide temperature range from 10 K to 250 K for the normalized intensity of magnetic peaks and lattice constant $c$ suggests a strong coupling between the spin and lattice ("magneto-elastic" coupling) in Cr$_2$GaN. This is extremely alike the typical characteristics of Mn-based MAX material Mn$_2$CaC and MnNiGe, where the magnetostriction effect is accompanied with the coupling among spin, lattice and electrical transport properties \ucite{7,8,17}.

Recent research has suggested the potential for the practical application of the MAX phases. The challenges in the immediate future are to further explore and characterize the MAX phases reported so far and to make further advance in boosting their industrial application. Here, the presented experimental results show that Cr$_2$GaN exhibits a magnetic transition at 170 K, below which all Cr atoms are aligned with sinusoidal modulated magnetic structure. Accompanied with this magnetic transition, drastic change in the crystal structure of the material take places, where 0.4\% contraction in the out-of-plane $c$ axis with increasing temperature from 10 K to 170 K indicates that it is a considerably large effect compared to what has commonly been observed in magnetostriction systems. The reported magneto-elastic coupling in Cr$_2$GaN is much like that in Mn-based MAX phase Mn$_2$GaC, in which the anisotropic structural changes are argued to be driven by magnetism \ucite{8}. For Cr$_2$GaN, by simply stretching the lattice in the $c$ direction, the Cr-Cr interplayer distance is altered and thereby weakening magnetic coupling across the Ga layer, thus it becomes available to be capable of shifting the occurrence of magnetic transitions, as reported by doping and pressure effects in this compound \ucite{12} . Due to the magneto-elastic coupling interaction, it would be interesting for further studies to explore possible anisotropic response of the lattice and electrical conductivity to magnetic fields applied parallel and perpendicular to the $c$ axis. Therefore, the single crystals of Cr$_2$GaN are highly desired for transport experiments to explore possible magnetoelectric or magnetocaloric applications.

Moreover, Cr$_2$GaN may be a candidate for topological magnetic texture or electronic states under its complex and tunable magnetism. To date, many novel magnetic topological materials, such as Co$_2$MnGa \ucite{18-20}, Co$_2$MnAl \ucite{21,22}, Co$_2$Ti$X$ ($X$ = Si, Ge and Sn) \ucite{23}, $X$Co$_2$$Z$ ($X$ = IVB or VB; $Z$ = IV A or IIIA) \ucite{24}, Co$_3$Sn$_2$S$_2$ \ucite{25,26} and SrRuO$_3$ \ucite{27,28}, have been identified by combining topology of electronic band structures with magnetic configurations. These novel materials exhibit rich topological states due to strong spin-orbit coupling or magneto-elastic coupling, referred such as magnetic topological insulators, Weyl and Dirac magnetic semimetals, axionic insulators or higher-order topological phases of matter, etc \ucite{29-36}. For Cr$_2$GaN as a 211-type compound similar to those Co-based systems, if the topological states exist, it will be very interesting to tune the topological states by its crystal structure continuously regulated by magnetism upon decreasing temperature.

In conclusion, the neutron powder diffraction and Rietveld refinement are carried out to determine the magnetic structure of Cr$_2$GaN, which is described as an incommensurate sinusoidal modulated structure with propagating vector  ${k}$=(0.365, 0.365, 0). Temperature-dependent diffraction patterns manifests no structural transition occurs, but a $c$-axis lattice stretching below the magnetic phase transition temperature $T_M\approx$ 170 K, which is also overlapped with the magnetic order parameter. The reported magneto-elastic coupling interplay in Cr$_2$GaN will provide us a platform for further explore the possible magnetoelectric, magnetostrictive, magnetocaloric effects and magnetic topological states in the MAX compounds.

\nocite{*}

\section*{References}

\vspace*{-0.8\baselineskip}\frenchspacing

\hskip 7pt {\footnotesize

\REF{[1]} Barsoum M W 2000 {\it Prog. Solid State Chem.} {\bf 28} 201
\REF{[2]} Eklund P \etal 2010 {\it Thin Solid Films} {\bf 518} 1851
\REF{[3]} Sun Z M 2011 {\it Int. Mater. Rev.} {\bf 56} 143
\REF{[4]} Barsoum M W \etal 2001 {\it Am. Sci.} {\bf 89} 334
\REF{[5]} Barsoum M W \etal 1996 {\it J. Am. Ceram. Soc.} {\bf 79} 1953
\REF{[6]} Ingason A S, Dahlqvist M and Rosen J 2016 {\it J. Phys.: Condens. Matter} {\bf 28} 433003
\REF{[7]} Novoselova I P \etal 2018 {\it Sci. Rep.} {\bf 8} 2637
\REF{[8]} Dahlqvist M \etal 2016 {\it Phys. Rev. B} {\bf 93} 014410
\REF{[9]} Boucher R, Berger O and Leyens C 2016 {\it Surf. Eng.} {\bf 32} 172
\REF{[10]} Salikhov R \etal 2015 {\it Mater. Res. Lett.} {\bf 3} 156
\REF{[11]} Liu Z, Waki T, Tabata Y and Nakamura H 2014  {\it Phys. Rev. B} {\bf 89} 054435
\REF{[12]} Li Y F, Liu J Z, Liu W H, Zhu X Y and Wen H H 2015 {\it Philos. Mag.} {\bf 95} 2831
\REF{[13]} Liu Z, Waki T, Tabata Y, Yuge K, Nakamura H and Watanabe I 2013 {\it Phys. Rev. B} {\bf 88} 134401
\REF{[14]} Rietveld H M 1969 {\it J. Appl. Crystallogr.} {\bf 2}
\REF{[15]} Rodr\'{i}guez-Carvajal J 1993 {\it Physica B: Condensed Matter} {\bf 192} 55
\REF{[16]} Wills A S 2000 {\it Physica B: Condensed Matter} {\bf 276} 680
\REF{[17]} Ren Q Y, Hutchison W D, Wang J L, Studer A J and Campbell S J 2018 {\it Chem. Mater.} {\bf 30} 1324
\REF{[18]} Akito S \etal 2018 {\it Nat. Phys.} {\bf 14} 1119
\REF{[19]} Manna K \etal 2018 {\it Phys. Rev. X} {\bf 8} 041045
\REF{[20]} Guin S N \etal 2019 {\it NPG Asia Mater.} {\bf 11} 16
\REF{[21]} Vilanova Vidal E, Stryganyuk G, Schneider H, Felser C and Jakob G 2011 {\it Appl. Phys. Lett.} {\bf 99} 132509
\REF{[22]} Li P G \etal 2020 {\it Nat. Commun.} {\bf 11} 3476
\REF{[23]} Chang G Q \etal 2016 {\it Sci. Rep.} {\bf 6} 38839
\REF{[24]} Wang Z J \etal 2016 {\it Phys. Rev. Lett.} {\bf 117} 236401
\REF{[25]} Liu C \etal 2020 {\it Sci. China-Phys. Mech. Astron.} {\bf 64} 217062
\REF{[26]} Liu C \etal 2021 {\it Sci. China-Phys. Mech. Astron.} {\bf 64} 257511
\REF{[27]} Itoh S \etal 2016 {\it Nat. Commun.} {\bf 7} 11788
\REF{[28]} Jenni K \etal 2019 {\it Phys. Rev. Lett.} {\bf 123} 017202
\REF{[29]} Xu Y F \etal 2020 {\it Nature} {\bf 586} 702
\REF{[30]} Wang P Y, Ge J, Li J H, Liu Y Z, Xu Y and Wang J 2021 {\it The Innovation} {\bf 2} 100098
\REF{[31]} Bernevig B A, Felser C and Beidenkopf H 2022 {\it Nature} {\bf 603} 41
\REF{[32]} Wang H \etal 2020 {\it Sci. China-Phys. Mech. Astron.} {\bf 63} 287411
\REF{[33]} Xie X C 2021 {\it Sci. China-Phys. Mech. Astron.} {\bf 64} 217061
\REF{[34]} Yan B H 2021 {\it Sci. China-Phys. Mech. Astron.} {\bf 64} 217063
\REF{[35]} Cai Z \etal 2020 {\it Phys. Rev. B} {\bf 103} 134408
\REF{[36]} Li J, Feng J S, Wang P S, Kan E J and Xiang H J 2021 {\it Sci. China-Phys. Mech. Astron.} {\bf 64} 286811
}

\end{document}